\documentstyle[pra,aps,multicol]{revtex}

\begin{document}
\title{Security of classical noise-based cryptography}
\author{A. Tomita$\dagger $ and O. Hirota$\ddagger $}
\address{$\dagger $Fundamental Research Laboratories, NEC Corporation, Tsukuba\\
305-8501, Japan. \\
$\ddagger $Research Center for Quantum Communication, Tamagawa University,\\
Machida, Tokyo 194-8610, Japan}
\date{\today }
\maketitle

\begin{abstract}
We examine the security of a protocol on cryptographic key distribution
proposed by Yuen and Kim (1998 {\it Phys. Lett.} {\it A }{\bf 241} 135).
Theoretical and experimental analysis shows that, even if the eavesdropper
could receive more photons than the legitimate receiver, secure key
distribution is possible as long as the signal-to-noise-ratio of the
eavesdropper does not exceed eight times (9 dB) that of the receiver. Secure
key distribution was demonstrated using conventional fiber optics. The
secure key transmission rate in the experiment was estimated to be 2 Mb/s at
its maximum (0.04 bit per sender's bit.) The present protocol has advantages
over other quantum key distribution protocols in that it is more efficient
and more easily implemented, but careful design and management are necessary
to ensure the security of the cryptosystem.
\end{abstract}



\section{Introduction}

\noindent Cryptography is used to transmit a message from a sender (referred
to as Alice) to a receiver (Bob) without leaking useful information to
others. It has been proved that a message can be transmitted securely if it
is coded and decoded by a sequence of random bits (key) whose length is
equal to that of the message. The problem of secure transmission is then
reduced to that of generating a secret key shared by Alice and Bob.
Classical cryptosystems rely on computational complexity and may be broken
by an effective algorithm or a powerful computer. Quantum key distribution
(QKD) protocols, in contrast, provide an unconditionally secure key, the
security of which is inherent in the laws of quantum mechanics. This
remarkable advantage of QKD protocols has been attracting increasing
research interest\cite{BB84,Wiesner,Ekert 91,BBM92,B92} since the proposal
by Bennett and Brassard\cite{BB84}. Although QKD has been demonstrated in
over-20-km-fiber communication channels\cite{Marand 95,Muller 97,Merolla
99,Bourennane 99,Hughes 99}, its application in practical communication
systems is not straightforward. The QKD protocols require single-photon
transmission to guarantee the security, and thus are vulnerable to loss and
noise inherent in actual transmission channels. Optical amplifiers will not
solve this problem, because the noise of the optical amplifiers inevitably
destroys the quantum correlation. Single-photon transmission requires the
use of complicated and inefficient photon counting techniques instead of
conventional analog detection, besides a truly practical single-photon
source is not yet available. The QKD protocols are therefore not fully
compatible with the current optical fiber communication systems. A secure
key distribution protocol compatible with the current systems is desirable.
This would be a protocol that uses more than one photon and allows optical
amplifiers to be used. Such a protocol would be based on coherent state
photons, or classical light.

Maurer\cite{Maurer 93} has shown perfect cryptographic security can be
obtained in a classical noisy channel with the help of a noiseless feedback
channel. Yuen and Kim\cite{Yuen 98} examined the principles underlying the
QKD protocol with two non-orthogonal quantum states (B92 protocol\cite{B92}%
.) The security of the B92 protocol relies on two facts\cite{Ekert 94}: (i)
an eavesdropper (Eve) cannot accurately determine the value of each
transmitted bit ( i.e., no efficient opaque eavesdropping.) (ii) Eve cannot
closely correlate Bob's measurement results with her own ( i.e., no
efficient translucent eavesdropping.) Yuen and Kim\cite{Yuen 98} pointed out
that these two conditions can be satisfied in a classical transmission
system, where the detectors of Bob and Eve are under independent additive
noise and show a small signal-to-noise-ratio (SNR.) They proposed a
classical noise-based protocol for key distribution (referred here to as the
YK protocol.) The YK protocol working with classical light, would have
advantages in practical implementations. Proving the security in YK protocol
is, however, subtler than in QKD protocols. Since many photons are
transmitted to carry one-bit information, the conditions specified above
will not be satisfied if the SNR of Eve's detection is sufficiently high.
Eve's SNR can be increased by using low-noise detection equipment, or simply
by moving closer to Alice than Bob (because of the fiber loss.) The original
analysis of the security of YK protocol assumed the same SNR for Bob and Eve
\cite{Yuen 98}. For practical implementations, it is important to determine
the design rules of Eve's SNR and Bob's.

In this article we quantitatively examine the security of the YK protocol,
and show that the YK protocol is secure, even if Eve's SNR is 9 dB better
than Bob's. We also show experimental results that demonstrate secure key
distribution against translucent attack. Section 2 provides the condition
for secure key distribution in terms of the secure key distribution rate.
Section 3 describes the experiment on the YK protocol using conventional
fiber optics. Section 4 discusses the implementation issues.

\bigskip

\section{Theory}

We first define the secure key distribution rate. Suppose Alice transmits an
equally probable binary string to Bob. Shannon information between Alice and
Bob is expressed by 
\begin{equation}
I_{AB}=1+e_{B}\log _{2}e_{B}+\left( 1-e_{B}\right) \log _{2}\left(
1-e_{B}\right) ,  \label{IAB}
\end{equation}
where $e_{B}$ denotes the error rate of Bob's decision. Because of the
decision errors, Bob has the information in only $n_{sift}I_{AB}$ of the $%
n_{sift}$ sifted bits. Alice and Bob exchange redundant information over the
public channel in order to obtain the reconciled key. This procedure is
called {\it error correction}, the best known practical protocol for which
was given by Brassard and Salvail\cite{BS93}. For successful error
correction with Brassard-Salvail's protocol, the error rate should be less
than 0.15. To establish a secret key, Alice and Bob use {\it privacy
amplification}\cite{BBCM95}, random hashing of the reconciled key into a
shorter key. If they shorten the reconciled key of length $n_{rec}$ by the
fraction $\tau $ and sacrifice $n_{S}$ bits as a safety parameter, Eves's
Shannon information on the final key of length $\tau n_{rec}-n_{S}$ is
bounded by 
\begin{equation}
I_{E}\leq \frac{2^{-n_{S}}}{\ln 2}.
\end{equation}
The fraction $\tau $ is given by 
\begin{equation}
\tau =1+\left( 1/n_{rec}\right) \log P_{C},  \label{def-t}
\end{equation}
where the collision probability $P_{C}\left( X\right) $ of $X$ is defined as
follows: Let $X$ be a random variable with an alphabet ${\cal X}$ and
distribution $P_{X}$. The collision probability is the probability that $X$
takes the same value twice in two independent experiments, that is, $%
P_{C}\left( X\right) =\sum_{x\in {\cal X}}P_{X}\left( x\right) ^{2}$. The
logarithm of the collision probability thus refers to Eve's information on
the key. The collision probability can be expressed by the probability $p(k)$
that $k$ is the $i$-th signal of Bob's string and the joint probability $%
p(k,l)$ that $k$ is the $i$-th signal of Bob's string and $l$ is the $i$-th
signal of Eve's string. We have the following formula\cite{Lutkenhaus 96}
for the fraction $\tau $: 
\begin{equation}
\tau =1+\log _{2}\left[ 
\mathrel{\mathop{\sum }\limits_{k,l}}%
\frac{p(k,l)^{2}}{p(k)}\right] .  \label{fraction}
\end{equation}
According to Bru\ss\ and L\"{u}tkenhaus\cite{BL99}, Bob can generate secure
bits from his sifted bits at the rate $R$ of 
\begin{equation}
R=I_{AB}-\left( {1}-e_{B}\right) \tau -e_{B}.  \label{rate}
\end{equation}
We refer this rate $R$ as the secure key distribution rate, and for secure
key distribution its value be positive. The actual key generation rate is
further reduced by multiplying the generation rate of the sifted key. In the
following part of this section, we derive the conditions under which $R$ is
positive.

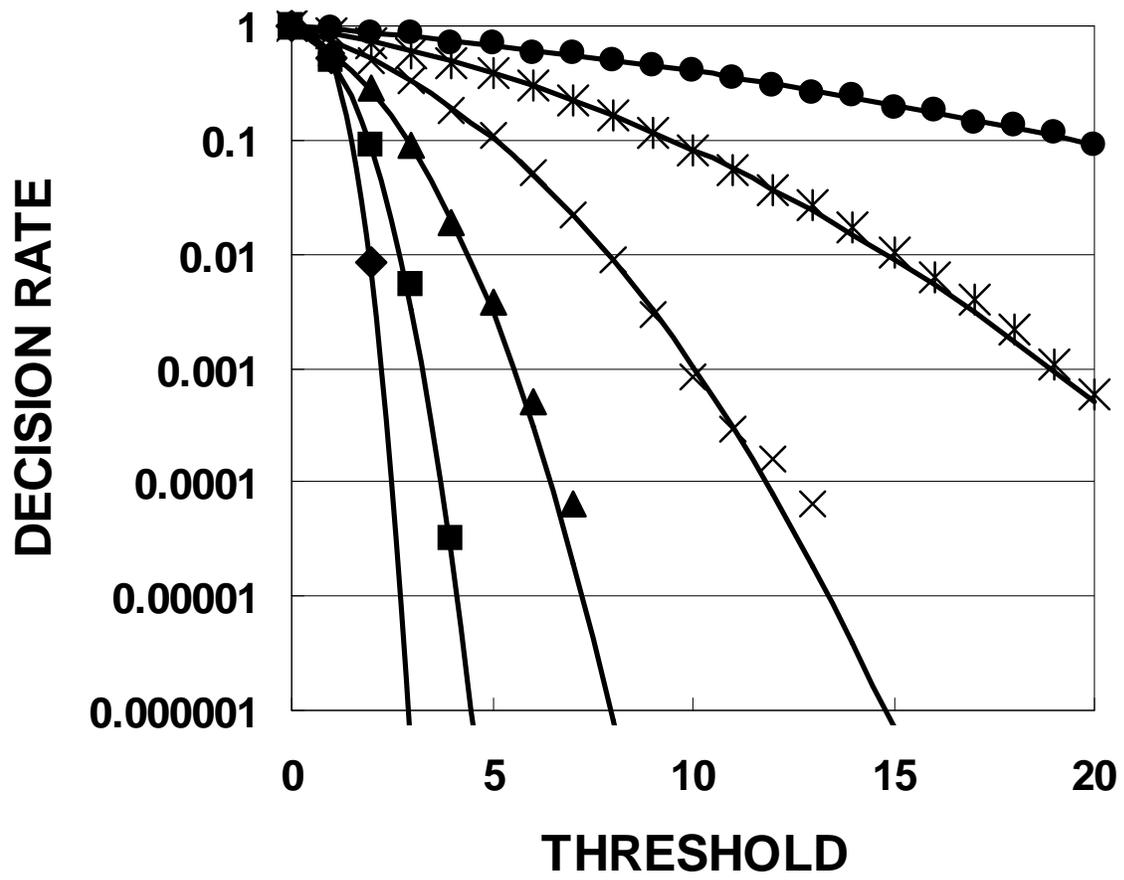
\begin{figure}[tbp]
\caption{Relation between threshold and the decision rate obtained at
various signal-to-noise-ratios: diamonds 7.8 dB, squares 2.65 dB, triangles
-3.28 dB, crosses -9.25 dB, stars -15.1 dB, and circles -21.4 dB. Lines were
calculated assuming white Gaussian noise. Symbols show values obtained in
the experiment.}
\label{figone}
\end{figure}

In the YK protocol\cite{Yuen 98}, the bit values (``0'' and ``1'') are
encoded so as to make the probability distribution of the received signal
symmetric. Alice sends encoded bits on a weak classical light. Signal $%
s_{0}(t)=S\phi (t)$ is transmitted for ``0'', and $s_{1}(t)=-S\phi (t)$ is
transmitted for ``1'', where $\int_{T}\phi (t)dt=1$. We here measure the
signal value as the voltage on the load resistance $R_{load}$ of a
photodiode. Mean signal voltage $S$ is defined by $S^{2}=%
\int_{T}s_{i}^{2}(t)dt$, and $S^{2}/R_{load}$ represents the signal energy
over the duration $T$ (signal energy per bit.) The output $r(t)$ of the
detector contains the noise $n(t)$, so $r(t)=s_{i}(t)+n(t)$. If the noise is
white Gaussian noise with spectral density $\sigma ^{2}$, the probability
distribution of the detected signal $V$ is expressed by

\begin{equation}
P(V)=%
{\left( 1/\sqrt{2\pi }\right) \exp \left[ -\left( V-S\right) ^{2}/\left( 2\sigma ^{2}\right) \right] \qquad \left( for\ ``0"\right)  \atopwithdelims\{. \left( 1/\sqrt{2\pi }\right) \exp \left[ -\left( V+S\right) ^{2}/\left( 2\sigma ^{2}\right) \right] \qquad \left( for\ ``1"\right) }%
,  \label{PBD}
\end{equation}
where the signal is averaged over the duration $T$ as $V=\int_{T}r(t)dt$.
The SNR $\beta ^{2}$ in this system is define by $\beta =S/\sigma $. In a
conventional decision scheme the bit values are determined to be ``0'' if $%
V>0$ and ``1'' if $V<0$. Decision errors will occur at the rate of $Q\left(
\beta \right) $, where $Q$ is the scaled complementary error function
defined by 
\begin{equation}
Q(x)={\frac{1}{\sqrt{2\pi }}}\int_{x}^{\infty }\exp \left( -y^{2}/2\right)
dy.  \label{Qfunc}
\end{equation}
We set a threshold $V_{th}=mS\,(m>1)$ to make a decision:``0'' if $V>$ $%
V_{th}$ and ``1'' if $V<-V_{th}$, but leave {\it inconclusive} if $%
-V_{th}\leq V\leq V_{th}$. The probability of making a decision is given by
the following decision rate: 
\begin{equation}
F_{+}=Q\left( \left( m+1\right) \beta \right) +Q\left( \left( m-1\right)
\beta \right) ,  \label{decision}
\end{equation}
and the error rate is 
\begin{equation}
e=\frac{Q\left( \left( m+1\right) \beta \right) }{F_{+}}.  \label{error}
\end{equation}
The sifted key is generated from the raw bit string by the Bob's decision.
The decision rate $F_{+}$ thus refers to the generation rate of the sifted
key. As seen in Eqs. (\ref{decision}) and (\ref{error}), the decision rate $%
F_{+}$ and the error rate $e$\ are determined by the values of the SNR and
the threshold. As described below, this error rate determines the joint
probabilities $p(k,l)$ and therefore the secure key distribution rate. The
system is thus fully characterized by the SNR and the threshold. The error
rate can be reduced by increasing the threshold, but, a high threshold will
also reduce the decision rate. Since, as we can see by comparing Fig. \ref
{figone} and Fig. \ref{figtwo}, the decision rate decrease faster than the
error rate, the threshold value should not be set too high.

As in the B92 protocol\cite{B92}, the inconclusive results play a essential
role in guaranteeing the security of the key distribution. A finite
threshold value of Bob enables him to make accurate decisions on his sifted
key at a cost of the generation rate. Eve, on the other hand, should make a
decision with zero threshold in order to obtain conclusive results for all
the transmitted bits. If Eve uses a finite threshold in her decision, she
will obtain the inconclusive results on the sifted bits. The assumption of
independent noise prevents Eve from predicting which bit Bob will obtain a
conclusive result. Eve can acquire no information from these inconclusive
bits. Since Eve's error rate $e_{E}$ is less than 1/2, she will obtain more
information by making a decision with zero threshold. Therefore, Bob can
make more accurate decisions on the sifted key bits than Eve can. That is,
Bob has more information than Eve, and can distill secure key bits with
Alice.

Now we will examine the conditions for security against eavesdropping. We
here consider only two simple kind of eavesdropping, translucent attack and
opaque attack. A translucent attack can be made by simply putting a beam
splitter in the transmission channel. The translucent attack to the YK
protocol, in contrast to those to the QKD protocols, will not change the
state of the transmitted light. The probability distribution of Bob's bits
is the same as that of Alice's,  $p(0)=p(1)=1/2$, because after error
correction Alice and Bob share completely correlated results. The joint
probabilities $p(k,l)$ are $p(0,0)=p(1,1)=(1-e_{E})/2$ and $%
p(0,1)=p(1,0)=e_{E}/2$. The fraction $\tau $ is calculated from Eq. (\ref
{fraction}) as 
\begin{equation}
\tau =1+\log _{2}\left( {1}-2e_{E}+2e_{E}^{2}\right) .  \label{trans-t}
\end{equation}
The secure key distribution rate can be estimated by using Eqs. (\ref{IAB}),
(\ref{rate}), and (\ref{trans-t}). Figure \ref{figthree} shows Eve's
required error rate as a function of Bob's. As Bob's error rate $e_{B}$
increases, Eve's error rate should be increased in order to obtain a
positive secure key distribution rate . For example, if Bob's error rate is
0.15, Eve should make errors at a rate greater than 0.27. This implies that
SNR of Eve's system should be less than 0.38 for white Gaussian noise. On
the other hand, Bob's SNR should be better than 0.057 to keep his error rate
smaller than 0.15 and his decision rate at 10$^{-3}$. The secure key
distribution is therefore possible even if Eve's SNR is six times (8 dB) as
large as Bob's. The tolerance of the SNR increases as Bob's error rate
decreases, and it reaches 10 dB for $e_{B}=0.01$.

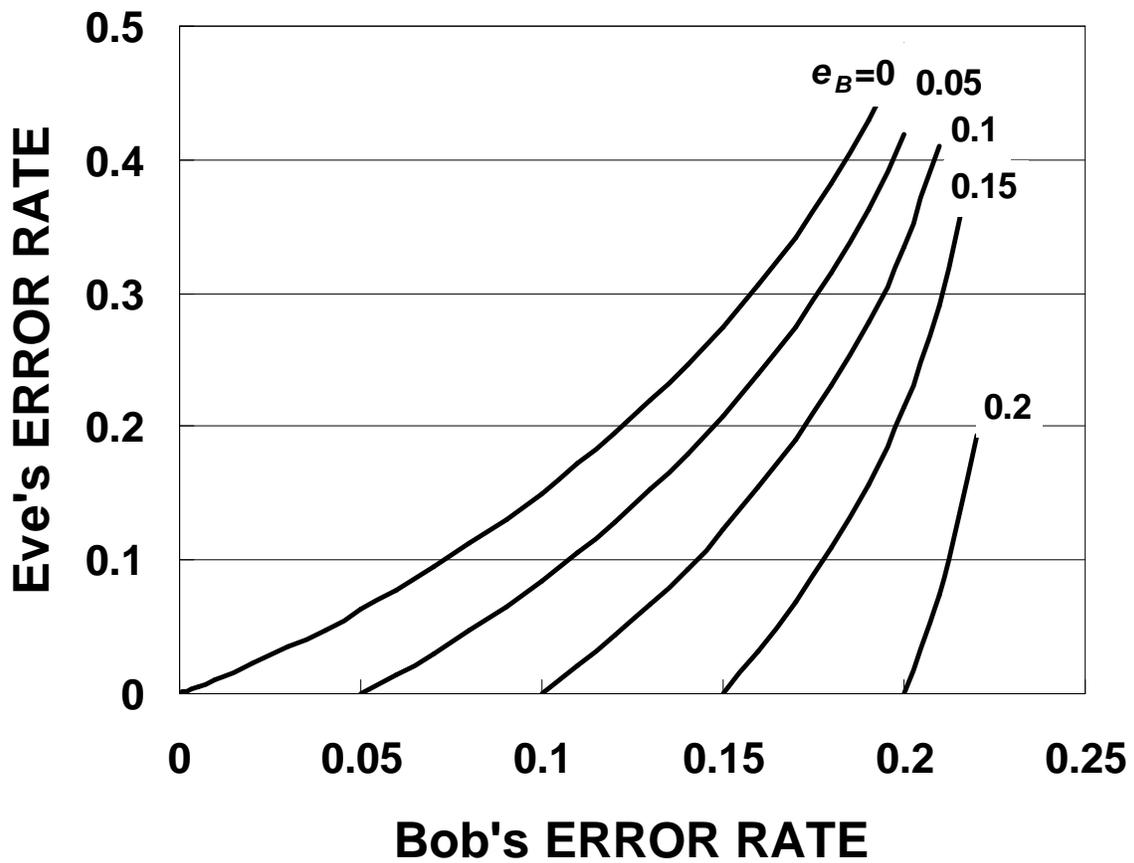
\begin{figure}[tbp]
\caption{The requirement for Eve's error rate as a function of Bob's error
rate to achieve secure key distribution against opaque attack. Lines are
calculated for the values of the Bob's error rate without eavesdropping.}
\label{figfour}
\end{figure}

In an opaque attack, Eve receives all the photons in $\eta n$ out of the $n$
bits sent by Alice. Then Eve sends the $\eta n$ bits to Bob according to her
decision. Eve never touches the rest of the bits ($\left( 1-\eta \right) n$
bits) and forwards them to Bob. To protect information from opaque attack,
Bob should determine his threshold according to the average signal intensity
of each bit. If he observes only the average intensity over many bits, Eve
can set a finite decision threshold to reduce her error rate and will then
obtain conclusive results for $\gamma \eta n$ bits ($\gamma <1.$) If she
sends only the conclusive results with signals $\gamma ^{-1}$ times as
intense as received, Bob will obtain the same long-time average signal
intensity he would if Eve did not intercept the photons. If Bob observes the
signal intensity of each bit, Eve must send every bit with the same
intensity as she receives it. Eve then should make a decision with zero
threshold, otherwise she will lose the information on the inconclusive
results. There is a trade-off for Eve on the fraction $\eta $: a large $\eta 
$ will increase Eve's information gain, but will also make her easily
detectable from the increase of Bob's error rate. Bob's error rate on the
unintercepted bits is $e_{B}$, but the error rate on the intercepted bits is 
$\left( 1-e_{E}\right) e_{B}+e_{E}(1-e_{B})$. The eavesdropping thus
increases Bob's error rate to 
\begin{equation}
e_{B}^{\prime }=\left( {1-\eta }\right) e_{B}+\eta \left[ \left(
1-e_{E}\right) e_{B}{+}e_{E}\left( 1-e_{B}\right) \right] .
\end{equation}
To calculate the secure key distribution rate by using Eqs. (\ref{IAB}), (%
\ref{rate}), and (\ref{trans-t}), we estimate the joint probabilities $%
p\left( k.l\right) .$ After the error correction, Bob has $(1-e_{B}^{\prime
})nF_{+}$ bits. The probability distribution i symmetric: $p(0)=p(1)=1/2.$
Eve obtains $(1-e_{E})\left( 1-e_{B}\right) \eta nF_{+}$ $+(1/2)(1-\eta
)(1-e_{B})nF_{+}$ correct results and $e_{E}e_{B}\eta nF_{+}+(1/2)(1-\eta
)(1-e_{B})nF_{+}$ incorrect results on Bob's bits. The joint probabilities
are obtained as 
\begin{eqnarray}
p\left( 1,1\right)  &=&\frac{\left[ (1-e_{E})\left( 1-e_{B}\right) \eta
+(1-e_{B})(1-\eta )/2\right] nF_{+}}{2\left( 1-e_{B}^{\prime }\right) nF_{+}}
\nonumber \\
&=&\frac{\left[ \left( 1-e_{E}\right) \eta +\left( 1-\eta \right) /2\right]
\left( 1-e_{B}\right) }{2\left( 1-e_{B}^{\prime }\right) }  \nonumber \\
p\left( 1,0\right)  &=&\frac{e_{E}e_{B}\eta nF_{+}+(1-e_{B})(1-\eta )nF_{+}/2%
}{2\left( 1-e_{B}^{\prime }\right) nF_{+}}  \label{jointP} \\
&=&\frac{e_{B}e_{E}\eta +(1/2)\left( 1-e_{B}\right) \left( 1-\eta \right) }{%
2\left( 1-e_{B}^{\prime }\right) }  \nonumber \\
p\left( 0,0\right)  &=&p\left( 1,1\right)   \nonumber \\
p\left( 0,1\right)  &=&p\left( 1,0\right) .  \nonumber
\end{eqnarray}
Figure \ref{figfour} shows the minimum required values of Eve's error rate
for secure key distribution $\left( R>0\right) $ as a function of Bob's
error rate $e_{B}^{\prime }$. Though Bob can observe only $e_{B}^{\prime }$
values, he can estimate $e_{B}$ from the SNR of his detection system. Eve
will be detected if $e_{B}^{\prime }\gg e_{B}$. The detection is easy if
Bob's error rate is much lower than Eve's. A high error rate for Bob may
hide Eve, but secure key distribution is possible even in this case. Suppose 
$e_{B}=0.1$ and $e_{B}^{\prime }=0.15$. As shown in Fig. \ref{figfour}, the
secure key distribution rate is positive if Eve's error rate is larger than
0.12. If the system is under white Gaussian noise, this condition on the
error rate is satisfied when Eve's SNR is smaller than 1.35 (1.3 dB.) Since
Bob's SNR should be better than 0.089 (-10.5 dB) to keep the decision rate
at 10$^{-3}$\ and $e_{B}=0.1$, the tolerance in SNR is 11.8 dB. This small
SNR for Eve implies that the signal should be sent on a weak light.
Increasing the light intensity reduces Eve's error rate, and makes the
secure key distribution impossible.

\section{Experiment}

In implementing the YK protocol, we should code the bit values in such a way
that the probability distribution of the received signals is symmetric . In
this experiment we used the unipolar Manchester code. This code represents
''1'' as a change from ON to OFF and ''0'' as a change from OFF to ON. It
can be decoded as follows: divide the incident light into two paths, one of
which is set one half of the pulse width longer than the other. Then take a
difference of the two light intensities by a balanced detector. The latter
half of the pulse slot yields a negative signal for ''1'' and a positive
signal for ''0''. Binary phase shift keying (BPSK) also yields a symmetric
distribution by homodyne detection, and would be more sensitive, but
unipolar Manchester code is easier to implement.

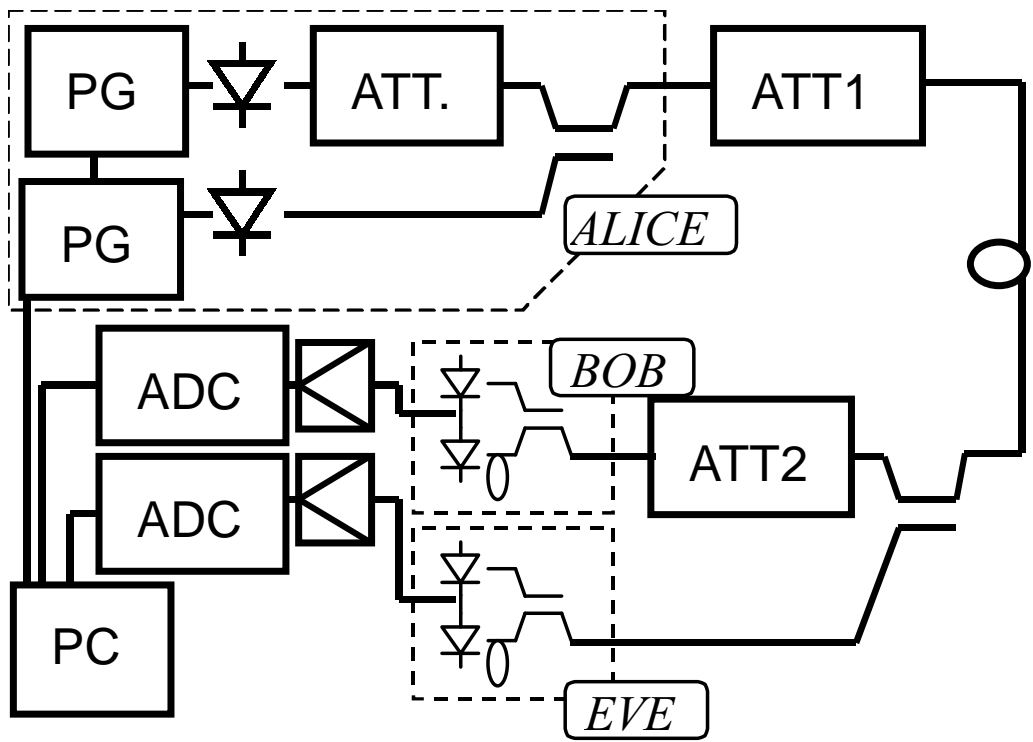
\begin{figure}[tbp]
\caption{Experimental set up for demonstration of the Yuen-Kim protocol. In
Alice's transmitter, pattern generators (PG) drive two lasers. ATT., ATT1
and ATT2 are attenuators. In the receivers of Bob and Eve, the light in one
of the divided path is delayed. Lights are detected by balanced detectors
made of two photodiodes. ADC: analog-digital converters. PC: personal
computer.}
\label{figfive}
\end{figure}

Figure \ref{figfive} shows the experimental setup. Two distributed feedback
(DFB) laser diodes (LDs) served as 1.3 $\mu $m light sources. A pattern
generator provided a signal pulse string to modulate one DFB LD (signal LD)
directly. The second pattern generator was synchronized to the first, and
provided an 8-ns pulse at the beginning of each pulse string. This pulse
modulated the other DFB LD (trigger LD) directly to generate a trigger light
pulse. The output of the signal LD was set weaker than that of the trigger
LD. The clock frequency in the present experiment was 25 MHz. Only a fixed
pattern of 101010$\cdots $ was transmitted. The coded signal light then
became a square wave with a duty of 50 \% and a pulse duration of 20 ns. We
sent strings of 30.8 kbits. The outputs of the two LDs were combined and
attenuated by an attenuator (ATT1.) To simulate the translucent attack by an
eavesdropper, we inserted a 50:50 divider. An attenuator \ (ATT2) was placed
in one arm of the divider to examine the SNR tolerance for the secure key
distribution. The signals of both outputs were detected by the receivers.
Each receiver consisted of a 50:50 divider, a fiber delay of a half pulse
width, and a balanced detector. The balanced detectors made of two
commercial InGaAs {\it pin} photodiodes loaded by 50 $\Omega $ resisters
were operated in analog mode. The catalog data (typical values) for the
quantum efficiency and the dark current \ of the photodiodes at 25 C were 90
\% and \ 5 nA. The photodiodes ware not cooled. The output signals of the
receivers were led to amplifiers ($G=40$ dB) and then to analog-digital
converters.

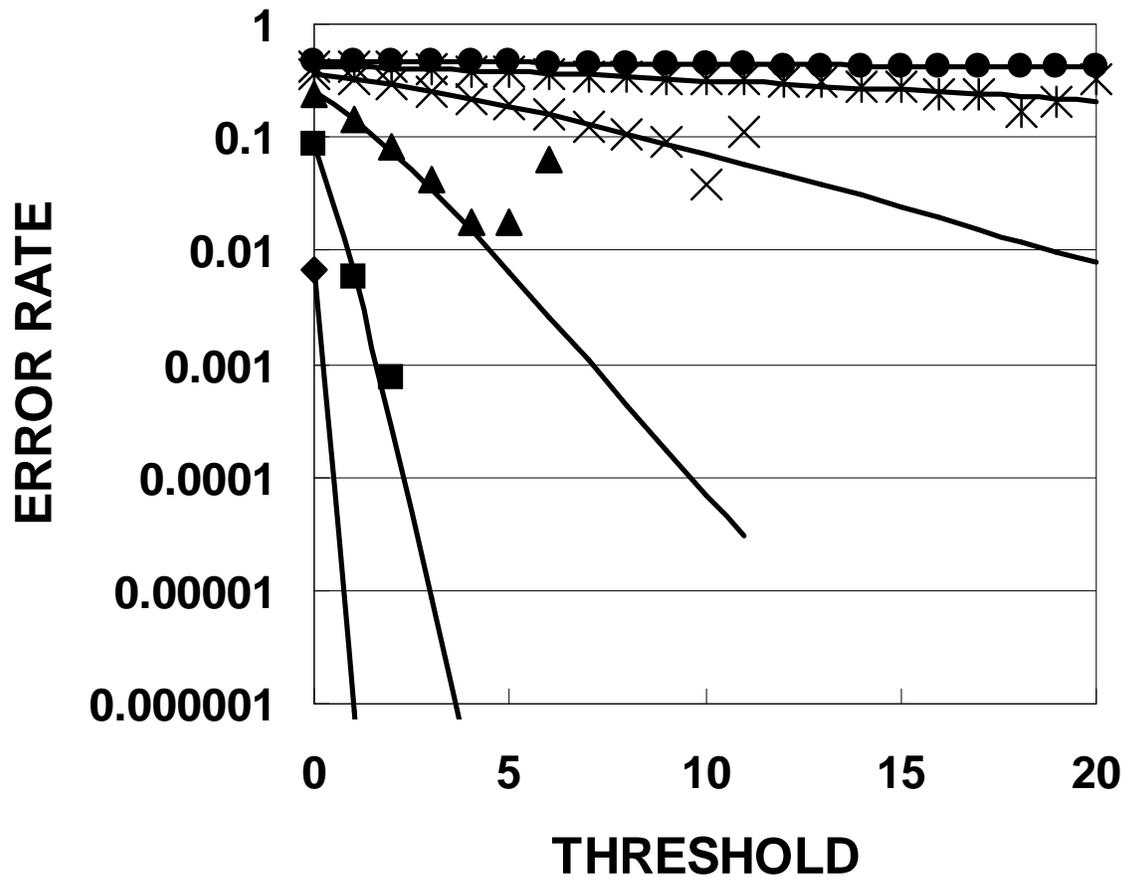
\begin{figure}[tbp]
\caption{Relation between threshold and the error rate obtained at various
signal-to-noise-ratios. The meanings of the symbols are the same as in Fig.
1.}
\label{figtwo}
\end{figure}

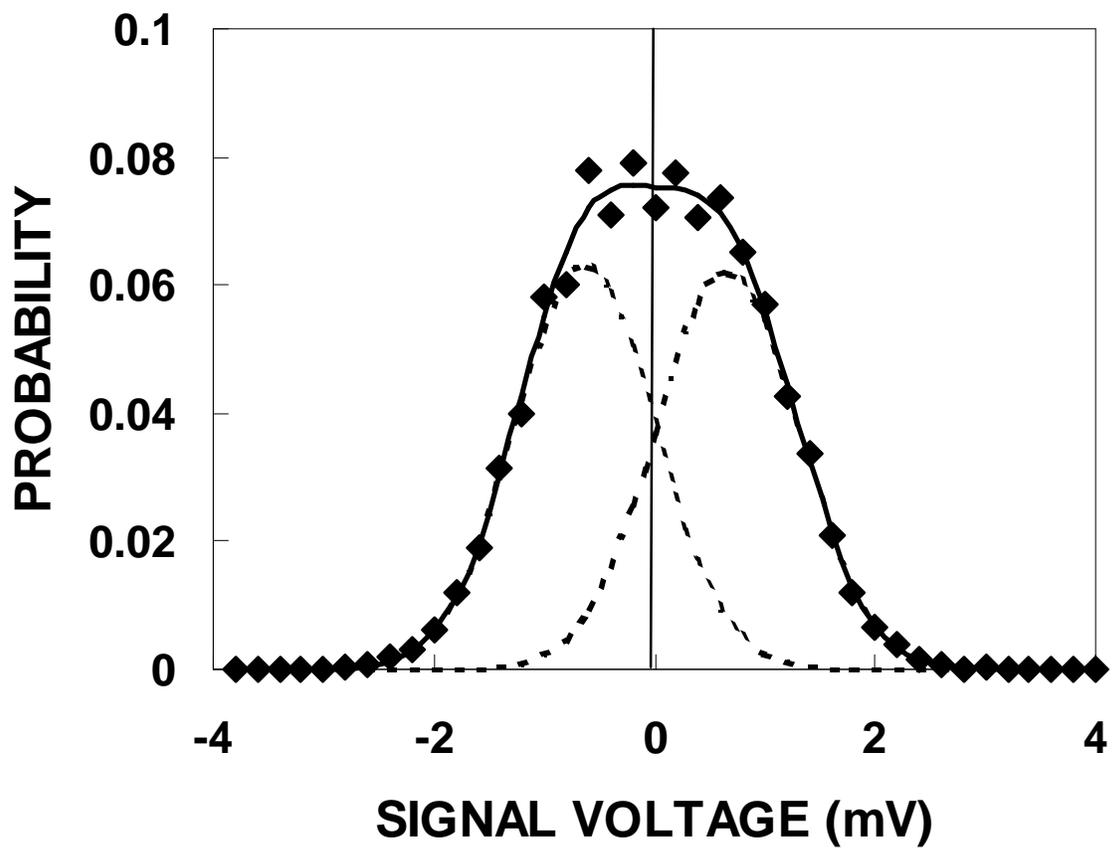
\begin{figure}[tbp]
\caption{A typical probability distribution of the output signal from the
amplifier. Diamonds denote experimental result, broken lines show the
Gaussians used for fitting, and the solid line shows the sum of those two
Gaussians.}
\label{figsix}
\end{figure}

Figure \ref{figsix} shows a typical probability distribution of the output
signal from the amplifier. It is well represented by the sum of two
Gaussians. The intensity of the optical signal was 0.380 $\mu $W (-34.2 dBm)
at the input port of the receiver, and the SNR of this signal was 1.0 (0
dB.) We averaged the output pulse over the duration (10 ns), and evaluated
the decision rate and error rate as a function of the SNR and the threshold.
The results are shown in Fig. \ref{figone} and Fig. \ref{figtwo}. The
experimental results agree well with the theory assuming white Gaussian
noise. These indicated that white Gaussian noise dominated the present
receiver sensitivity, and that the security analysis described in Sec. 2 can
be applied to the experiment. The number of the sifted bits became small
when the threshold value is high. We had less than 30 bits, if the decision
rate is less than 10$^{-3}$. This insufficient sample number caused the
error rate fluctuation observed for large $m$'s in Fig. \ref{figtwo}. For
SNR values up to 0 dB, the noise level was almost same as the dark noise
level, but for larger SNR, it increased with the signal intensity. Dark
current of the photodiodes was negligible compared to the thermal noise.
These indicates that, for SNR values up to 0 dB, the sensitivity of the
system was dominated by thermal noise, which is constant to the input photon
number.  As the intensity increased, the thermal noise was exceeded by shot
noise, which is proportional to the input photon number. SNR was
proportional to the square of the input power for weak signals, and tended
to be proportional to the input power as the signal intensity increased.

The error rate shown in Fig. \ref{figtwo} provides a criterion for key
distribution secure against opaque attack. A low error rate of 0.038 was
obtained for weak signals by setting the threshold at $m=10$, where the SNR
was -9.25 dB. The decisions were made at the rate of 0.0008, slightly lower
than 10$^{-3}$. This error rate was lower than the theoretical value of
0.072 because of the fluctuation described above. Using $e_{B}=0$ line in
Fig. \ref{figfour}, we conclude that the key distribution is secure if Eve's
error rate is larger than 0.1, where we use the theoretical value of the
error rate (0.072) for Bob. This condition is satisfied if Eve's SNR is 0
dB, because we obtained the error rate of 0.15 in the experiment. Bob's
advantage in SNR was thus greater than 9.25 dB. This advantage was almost
constant for large SNR signals.

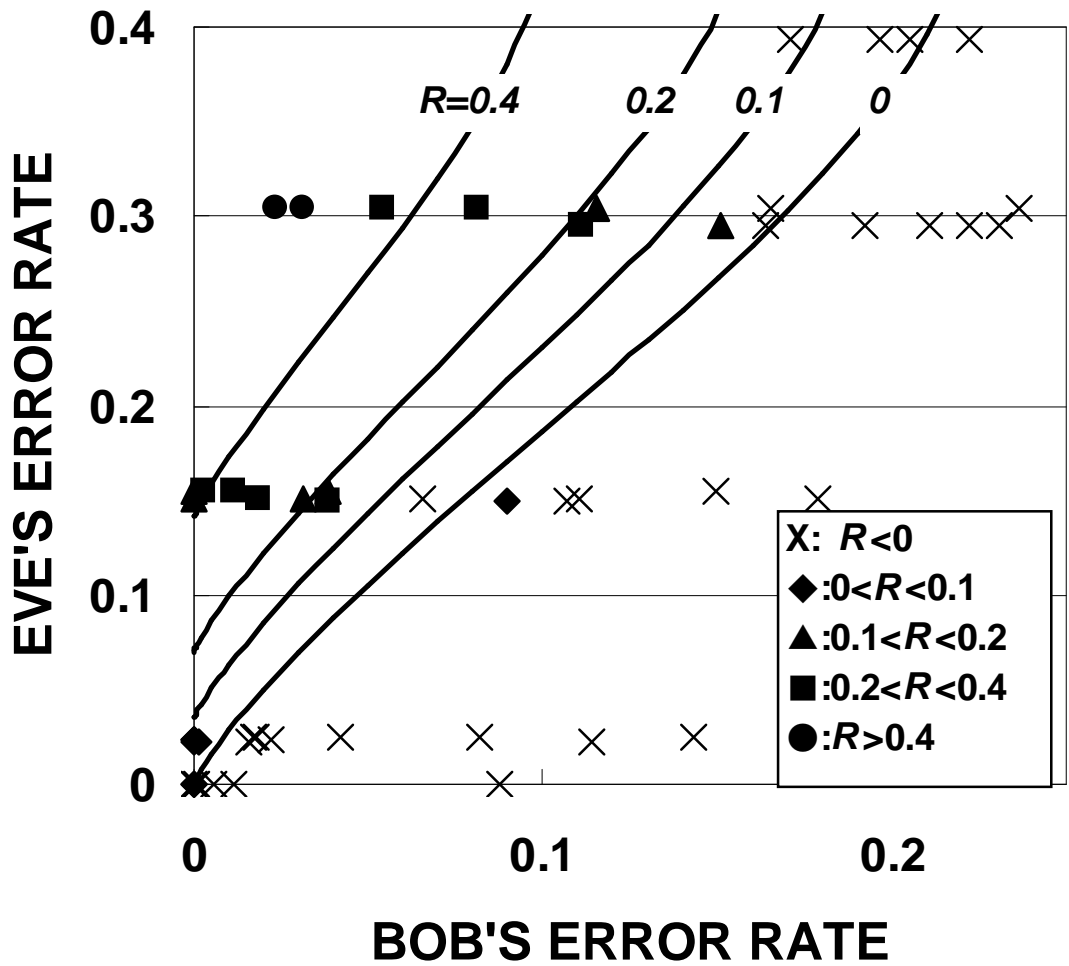
\begin{figure}[tbp]
\caption{The requirement for Eve's error rate as a function of Bob's error
rate to achieve secure key distribution against translucent attack. Lines
are calculated for the values of the secure key distribution rate $R=0$,
0.1, 0.2, and 0.4. Symbols represent the secure key distribution rate
obtained in the experiment. Crosses denote $R<0$, diamonds: $0<R<0.1$,
triangles: $0.1<R<0.2$, squares: $0.1<R<0.2$ and circles: $R>0.4$.}
\label{figthree}
\end{figure}

Security against the translucent attack was examined as follows. We assigned
one receiver that followed ATT2 as Bob, and the other receiver as Eve. ATT1
affected the SNRs of both Bob and Eve, whereas ATT2 determined the ratio of
the SNRs. The decisions in Bob were recorded with several values of the
threshold $m$, while the Eve's decisions were recorded with the threshold
fixed at zero. We measured the error rate $e_{B}$ and decision rate $F_{+}$
of Bob and the error rate $e_{E}$ of Eve. We estimated the joint
probabilities $p(0,0),$ $p(0,1)$, $p(1,0),$ and $p(1,1)$ from the bit data
about which Bob made correct decisions. Finally, we calculated the secure
key distribution rate $R$ by using Eqs. (\ref{IAB}), (\ref{fraction}), and (%
\ref{rate}). Figure \ref{figthree} shows the secure key distribution rate as
a function of the error rates of Eve and Bob. The symbols in Fig. \ref
{figthree} show the secure key distribution rates estimated from the
experiment. Experimental results agree well with theoretical results
(lines.) \ The secure key distribution was achieved if error rates of Eve
and Bob are in the region above the $R=0$ line in Fig. \ref{figthree}. The
decrease in Bob's SNR reduced the range of the signal intensity for secure
key distribution. Secure key distribution was impossible when Bob's SNR was
-9 dB smaller than  that of Eve. This result also agrees well with the
prediction. We obtained the largest actual secure key distribution rate $%
F_{+}R=0.04$ when the SNRs of both Bob and Eve were unity (0 dB) and Bob's
threshold was set to $m=2$. The observed error rates were 0.01 for Bob and
0.15 for Eve. The secure key distribution rate was $R=0.29$. Higher secure
key distribution rates were obtained by setting larger threshold values, but
the reduction in the decision rate decreased the product $F_{+}R$. Alice
transmitted signals at 50 Mb/s, so that the key transmission rate in the
present experiment was 2 Mb/s. This is a hundred times as fast as the key
transmission rate reported in the QKD experiments \cite{Marand 95,Hughes 99}%
. The transmission rate was limited only by the electric circuits. The
secure key would be transmitted at 400 Mb/s if a 10-Gb/s transmission
channel were used.

\section{Discussions}

This theoretical analysis has shown that the secure key distribution is
possible as long as the ratio of Bob's SNR to Eve's is better than -9 dB,
and the experimental results presented here confirmed it. A practical
cryptosystem should thus be designed to satisfy this condition. Eve may stay
much closer to Alice than Bob, and her signal may be larger than Bob's
because of \ the fiber loss. We estimate a limit of the transmission
distance in the following. It would be very difficult to use complicated
networks, where the path of a traffic is not fixed. We have to construct a
cryptosystem on a simple network or a point-point channel. Suppose, for
simplicity, we construct it on a point-to-point channel. SNR is proportional
to the square of the light intensity when the system sensitivity is limited
by thermal noise. Then Bob's advantage of 9 dB refers to the fiber length of
22.5 km using a lowest loss fiber (0.2 dB/km) and neglecting connection
loss. SNR is proportional to the light intensity in systems limited by shot
noise, and the fiber can be as long as 45 km in those systems. This values
would be increased assuming the translucent attack, because Eve would tap
the channel and receive small part of the signal.

Amplifiers can be used in YK protocol as long as the SNR permits. They will
improve the SNR by reducing the effect of the thermal noise, and therefore
will be useful when the system is limited by the thermal noise. In the shot
noise limit, even an ideal amplifier reduces the SNR by 3 dB. The use of
amplifiers is restricted by this degradation in the SNR. At most three
amplifiers are thus possible.

The above estimation assumed that Bob and Eve use the same detectors. Bob
should reduce system noise as possible to guarantee the security, by cooling
the receiver, for example. If he can suppress all the thermal noise, the SNR
of his system will be limited by shot noise, the standard quantum limit\cite
{YuenQCM98}. \ The mean photon number transmitted in this system should be
reduced to unity, because the SNR of 0 dB refers to mean photon number of
unity in the shot noise limited systems. The error rate can be reduced below
the standard quantum limit by optimum decision\cite{Helstrom,Sasaki96}. The
improvement will be apparent for small mean photon numbers. Security
analysis based on quantum detection theory as well as practical
implementation of the optimum decision are open for further study. It would
be noteworthy that the security analysis described in the present article
will provide a security criteria for the B92 protocols employing dim
coherent lights.

The YK protocol provides more efficient key distribution at higher bit rates
than do other QKD protocols, but it requires that the signal intensity be
controlled to keep Eve's SNR advantage smaller than 9 dB. This may be a
disadvantage compared to the QKD protocols like BB84, where the
unconditional security is proved if the photons are generated by a perfect
single photon source\cite{Mayers96,Mayers98}. However, it has been shown\cite
{BLMS00} that the Eve's advantage in SNR will limit the efficiency of the
protocol in a lossy channel. The SNR control would be also required in
actual BB84 systems.

\section{Conclusion}

Quantitative analysis of the security of the Yuen-Kim protocol shows that
the secure key distribution is possible even if the eavesdropper receives
signals with a signal-to-noise-ratio better than that with which the
legitimate receiver receives them. It has been shown that the
signal-to-noise-ratio of the legitimate receiver may be -9 dB smaller than
that of the eavesdropper. The results of an experiment using conventional
fiber optics agrees well with the analysis results. These results have
demonstrated a practical implementation of a secure key distribution
protected by the laws of physics. We think the YK protocol would be a
solution for practical cryptography systems.


\end{document}